\documentclass[aps,prl, groupedaddress,twocolumn]{revtex4-1}
\usepackage[pdftex]{graphicx}
\usepackage{subfigure}
\usepackage{sidecap}

\usepackage{amsmath}
\usepackage{hyperref}
\usepackage{color}

\newcommand{\gae}{\lower 2pt \hbox{$\, \buildrel {\scriptstyle >}\over {\scriptstyle \sim}\,$}}
\newcommand{\lae}{\lower 2pt \hbox{$\, \buildrel {\scriptstyle <}\over {\scriptstyle \sim}\,$}}

\begin{document}
\title{Enhancing the sensitivity of the LIGO gravitational wave detector by using squeezed states of light}

\author{J.~Aasi$^{1}$,
J.~Abadie$^{1}$,	
B.~P.~Abbott$^{1}$,	
R.~Abbott$^{1}$,
T.~D.~Abbott$^{9}$    	
M.~R.~Abernathy$^{1}$,	
C.~Adams$^{3}$,	
T.~Adams$^{36}$,
P.~Addesso$^{55}$,	
R.~X.~Adhikari$^{1}$
% ...
},	
\address{$^{1}$LIGO - California Institute of Technology, Pasadena, CA  91125, USA }
% ...
\address{$^{3}$LIGO - Livingston Observatory, Livingston, LA  70754, USA }
% ...
\address{$^{9}$Louisiana State University, Baton Rouge, LA  70803, USA }
% ...
\address{$^{36}$Cardiff University, Cardiff, CF24 3AA, United Kingdom }
% ...
\address{$^{55}$University of Sannio at Benevento, I-82100 Benevento, Italy and INFN (Sezione di Napoli), Italy }
% ...

\maketitle

\nopagebreak

\textbf{Nearly a century after Einstein first predicted the existence of gravitational waves, a global network of earth-based gravitational wave observatories~\cite{LIGO1992,LIGO2009,Virgo2008,GEO2010} is seeking to directly detect this faint radiation using precision laser interferometry. Photon shot noise, due to the quantum nature of light, imposes a fundamental limit on the attometer level sensitivity of the kilometer-scale Michelson interferometers deployed for this task.  Here we inject squeezed states to improve the performance of one of the detectors of the Laser Interferometer Gravitational-wave Observatory (LIGO) beyond the quantum noise limit, most notably in the frequency region down to 150 Hz, critically important for several astrophysical sources, with no deterioration of performance observed at any frequency. With the injection of squeezed states, this LIGO detector demonstrated the best broadband sensitivity to gravitational waves ever achieved, with important implications for observing the gravitational wave Universe with unprecedented sensitivity.}

A fundamental limit to the sensitivity of a Michelson interferometer with quasi-free mirrors comes from the quantum nature of light, which reveals itself through two fundamental mechanisms: photon counting noise (\emph{shot noise}), arising from statistical fluctuations in the arrival time of photons at the interferometer output; and \emph{radiation pressure noise}, which is the recoil of the mirrors due to the radiation pressure arising from quantum fluctuations in the photon flux. Both sources can be attributed to the quantum fluctuations of the electromagnetic vacuum field, or vacuum fluctuations, that enter the interferometer~\cite{Caves1980,Caves1981}.

\begin{figure*}[ht!]
\centering
\includegraphics[width=0.9\textwidth]{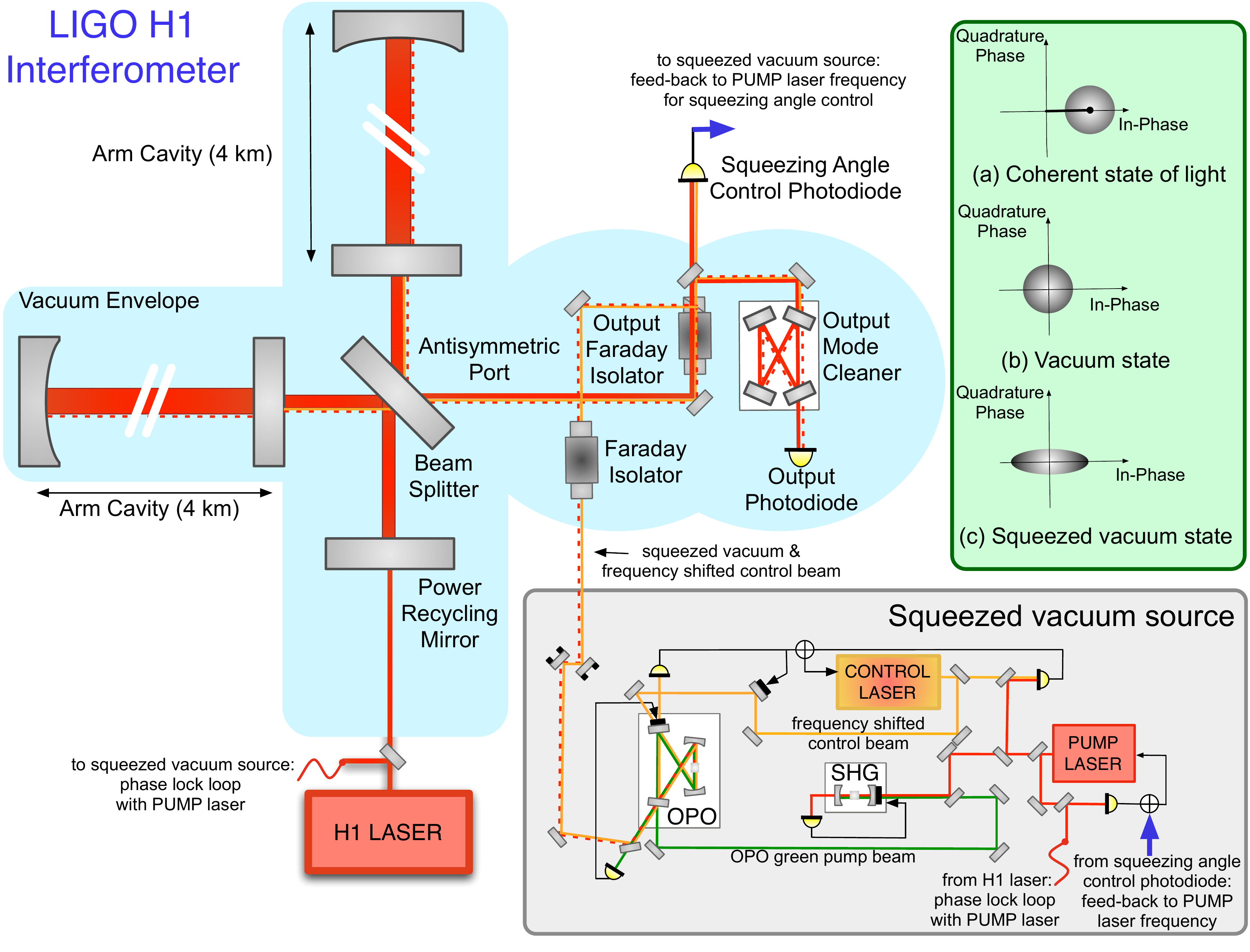}
\caption{ Simplified layout of the H1 interferometer with squeezed vacuum injection. The interferometer layout is described in the text, together with the main squeezer components (shown in the grey box). The green box shows a simplified representation of coherent states and squeezed states in the ``in-phase" and  ``quadrature phase" coordinates.
\label{fig:layout}}
\end{figure*}

An electromagnetic field can be described by two non-commuting conjugate operators that are associated with field amplitudes that oscillate out of phase with each other by $90^{\circ}$, labeled as ``in-phase" and ``quadrature phase"~\cite{GerryandKnight}. A coherent state of light (or vacuum, if the coherent amplitude is zero) has equal uncertainty in both quadratures, with the uncertainty product limited by the Heisenberg uncertainty principle.
For a squeezed state, the uncertainty in one quadrature is decreased relative to that of the coherent state (see green box in Fig.~\ref{fig:layout}). Note that the uncertainty in the orthogonal quadrature is correspondingly increased, always satisfying the Heisenberg inequality.

The vacuum fluctuations that limit the sensitivity of an interferometric gravitational wave detector enter through the antisymmetric port of the interferometer, mix with the signal field produced at the beamsplitter by a passing gravitational wave, and exit the antisymmetric port to create noise on the output photodetector. Caves~\cite{Caves1980,Caves1981} showed that replacing coherent vacuum fluctuations entering the antisymmetric port with correctly phased squeezed vacuum states decreases the ``in-phase"  quadrature uncertainty, and thus the shot noise, below the quantum limit. Shortly after, the first experiments showing squeezed light production through non-linear optical media achieved modest but important reductions in noise at high frequencies~\cite{Slusher1985}~\cite{Kimble1987}.
However, squeezing in the audiofrequency region relevant for gravitational wave detection and control schemes for locking the squeezed phase to that needed by the interferometer were not demonstrated until the last decade~\cite{Mckenzie2004}~\cite{Vahlbruch2006}~\cite{Vahlbruch2007}. Since then, squeezed vacuum has been used to enhance the sensitivity of a prototype interferometer~\cite{Goda2008}. The 600-m long GEO600 detector~\cite{GEOHFsqz2011} has deployed squeezing since 2010, achieving improved sensitivity at 700 Hz and above.

An important motivation for the experiment we present here was to extend the frequency range down to 150 Hz while testing squeezing at a noise level close to that required for Advanced LIGO~\cite{aLIGO2010}. This lower frequency region is critically important for the most promising astrophysical sources, such as coalescences of black hole and neutron star binary systems, but also poses a significant experimental challenge. Seismic motion is huge compared to the desired sensitivity, albeit at very low frequencies  $\lae 1$ Hz, and LIGO employs a very high performance isolation system to attenuate the seismic motion by several orders of magnitude. This uncovers a set of non-linear couplings which up-convert low frequency noise into the gravitational wave band. In the past, these processes have made it difficult for gravitational wave detectors to reach a shot noise limited sensitivity in their most sensitive band near 150 Hz. Any interactions between the interferometer and the outside world have to be kept at an absolute minimum. For instance, randomly scattered light reflecting back into the interferometer has to be managed at the level of $10^{-18}$~W.  Past experience has shown that measured sensitivities at higher frequencies are difficult to extrapolate to lower frequencies~\cite{LIGO2009}. For the first time, we employ squeezing to obtain a sensitivity improvement at a gravitational wave observatory in the critical frequency band between 150 Hz and 300 Hz.  Similarly important, we observed that no additional noise above background was added by our squeezed vacuum source, firmly establishing this quantum technology as an indispensable technique in the future of gravitational wave astronomy.

The experiment was carried out toward the end of 2011 on the LIGO detector at Hanford, Washington, known as ``H1''. The optical layout of the detector is shown in Fig.~\ref{fig:layout}. The interferometer light source (``H1 laser") is a Nd:YAG laser (1064 nm) stabilized in frequency and intensity. A beam splitter splits the light into the two arms of the Michelson, and Fabry-Perot cavities increase the phase sensitivity by bouncing the light $\approx 130$ times in each arm. The Michelson is operated on a dark fringe, thus most of the light is reflected from the interferometer back to the laser. A partially transmitting mirror between the laser and the beam splitter forms the power-recycling cavity, which increases the power incident on the beam splitter by a factor of 40.
In order to isolate them from terrestrial forces such as seismic noise, the power recycling mirror, the beam splitter, and the arm cavity mirrors are all suspended as pendula on 
vibration-isolated platforms.

A passing gravitational wave produces a differential change in the lengths of the arm cavities (generally, one arm gets shorter while the orthogonal arm gets longer), causing a signal field to appear at the antisymmetric port proportional to the wave amplitude.

For unperturbed arm length $L$, a gravitational wave of amplitude $h$ (in dimensionless units of strain) induces a differential change in arm length $\Delta L = h L$. For typical astrophysical sources from 10 to 100 Mpc away, such as the inspiral and merger of binary neutron stars or black holes, terrestrial detectors must measure strains at the level of $10^{-21}$ or smaller. 

A full description of this interferometer (and its sister interferometer in Livingston, LA)  can be found in Ref.~\cite{LIGO2009}.  A number of crucial modifications have been made since then that enable the implementation and testing of squeezing. In particular, the signal readout has been changed from a heterodyne to a homodyne system~\cite{DCtobin}, where we actively operated the Michelson interferometer with a small offset from a dark fringe to send about 30 mW of light to the antisymmetric port to act as the homodyne reference beam.  An output mode-cleaner (OMC in Fig.~\ref{fig:layout}) was also installed to prevent light in higher order optical modes and at different radio-frequency offsets from reaching the readout photodetector.
Moreover, the available laser power was increased from 10 W to 20 W.  This resulted in 15 W of light power reaching the interferometer, 600 W impinging on the beamsplitter and 40 kW stored in the interferometer arm cavities. These modifications resulted in a factor of 2 improvement in sensitivity above 500 Hz over the 2009 configuration.

The grey box of Fig.~\ref{fig:layout} shows a simplified schematic of the squeezed vacuum source. A sub-threshold optical parametric oscillator (OPO) in a bow-tie configuration~\cite{bowtieOPO2012}~\cite{sqzOPO2012} produces the squeezed vacuum state. Light at 532 nm pumps the OPO and produces squeezed vacuum at 1064 nm via parametric downconversion in a second-order nonlinear PPKTP crystal placed in the OPO cavity. The ``pump laser'' for the squeezed vacuum source is phase-locked to the ``H1 laser'' and it emits 1064 nm light which drives the second harmonic generator (SHG) to produce light at 532 nm. The ``control laser'' is phase-locked to the ``pump laser" to generate a frequency shifted coherent beam which enters the interferometer through the ``output Faraday isolator'', together with the squeezed vacuum. The interferometer reflects both fields back towards the OMC, and the beat between the frequency shifted coherent beam and the interferometer beam is detected by the  ``squeezing angle control photodiode" to control the phase of the squeezed vacuum field relative to the interferometer field~\cite{Vahlbruch2006}. The OMC filters out the frequency shifted coherent beam, while the squeezed vacuum reaches the ``output photodiode".

\begin{figure*}[ht!]
\centering
\includegraphics[width=0.9\textwidth]{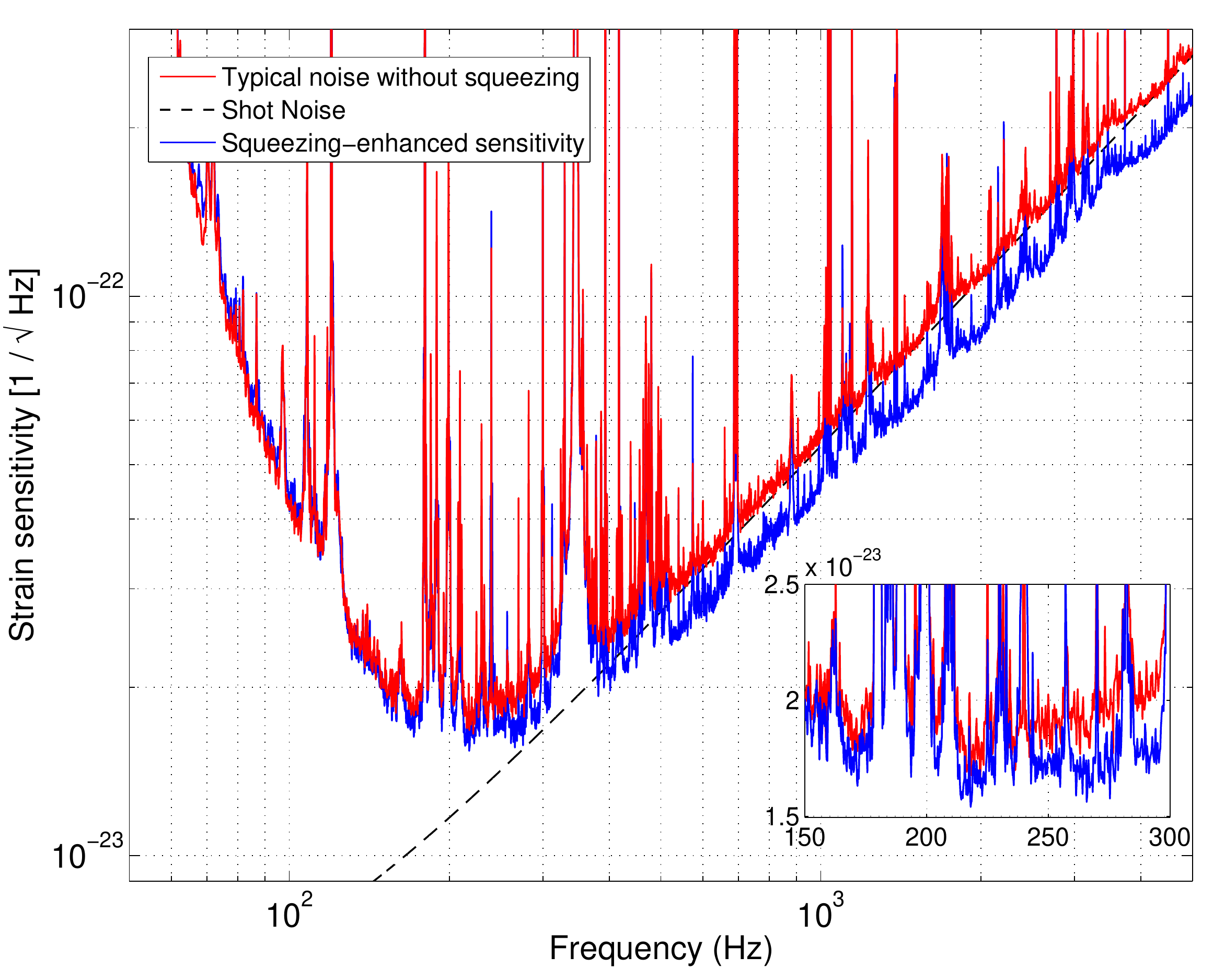}
\caption{Strain sensitivity of the H1 detector measured with and without squeezing injection. The improvement is up to 2.15 dB in the shot noise limited frequency band. Several effects cause the sharp lines visible in the spectra: mechanical resonances in the mirror suspensions, resonances of the internal mirror modes, power line harmonics, etc. As the broadband floor of the sensitivity is most relevant for gravitational wave detection, these lines are typically not too harmful. The inset magnifies the frequency region between 150 and 300 Hz, showing that the squeezing enhancement persists down to 150 Hz \label{fig:sens20W}}.
\end{figure*}  

During the experiment reported here, the LIGO H1 detector was configured as it was during its most sensitive scientific run S6~\cite{S62012} concluded in October 2010. Shot noise was the limiting noise source above 400 Hz and contributed significantly to the total noise down to 150 Hz~\cite{LIGO2009}. Radiation pressure noise was negligible, completely masked by other noise sources.  

The significantly improved sensitivity due to squeezing in this experiment is shown in Fig.~\ref{fig:sens20W}. The performance without squeezing shown by the red curve was comparable at high frequency to the best sensitivity H1 reached during S6. The blue curve shows the improvement in the sensitivity resulting from squeezing, with a 2.15 dB ($28\%$) reduction in the shot noise. This constitutes the best broadband sensitivity to gravitational waves ever achieved. To achieve the same improvement, a $64\%$ increase in the power stored in the arm cavities would have been necessary, but this power increase would be accompanied by the significant limitations of high power operation~\cite{aLIGO2010, ParametricInstabilities}. The measured improvement due to squeezing is well explained given the amount of squeezing injected into the interferometer and the total measured losses in the squeezed beam path, as we will detail later. A reduction in the total losses would therefore directly translate in a larger shot noise suppression. 

Equally important, the squeezed vacuum source did not introduce additional technical noise in any frequency band. This required paying particular attention in the design of the squeezed vacuum source to the control of scattered light. Scattered light is a serious issue in gravitational wave detectors operating near the quantum limit, since it not only limits the squeezing enhancement, but can degrade the interferometer sensitivity. Noise due to scattered light is also extremely hard to calculate a priori. Not only is the amount of scattering difficult to estimate, but the phase noise on the scattered light fields is also unknown. Understanding the impact of scattered light noise was one of the important motivations for our experiment. In particular, light leaving the interferometer is backscattered by the OPO and re-enters the interferometer, contaminating the main interferometer output. Because the squeezed vacuum source is mounted on an optical bench outside the vacuum system that houses the interferometer, large relative motions between the two are possible. To mitigate this problem, the H1 OPO is a traveling wave cavity designed to provide an intrinsic isolation to backscattering of more than 40 dB~\cite{bowtieOPO2012}.  Moreover, an additional Faraday isolator on an in-vacuum suspended isolation platform was installed in the injection path of the squeezed beam. With this arrangement, the backscatter noise due to linear phase variations was measured to be at least a factor 10 below the total noise in the critical region between 150 and 300 Hz.

In order to explain quantitatively the measured improvement in the LIGO H1 sensitivity, we studied the two main mechanisms that degrade squeezing: optical losses and phase noise.
Squeezed vacuum states are fragile; any optical loss, including imperfect mode matching, reduces the correlations imposed on the beam by the squeezed vacuum. Furthermore, fluctuations in the relative phase between the squeezed beam and the interferometer beam can degrade the quantum noise reduction, since deviation from the optimal phase projects the higher-noise orthogonal quadrature onto the measured quadrature. The measurement shown in Fig.~\ref{fig:sens20W} was obtained by injecting $10.3\pm0.2$ dB of squeezing, and the observed improvement in the shot noise limit is consistent with the measured losses and phase noise, as detailed below. 

Let us consider first the impact of optical losses. Given the normalized variance of the output mode $V_{\pm}$ for the elongated (+) and the squeezed (-) quadratures, respectively, the normalized variances $V^{'}_{\pm}$ for a given detection efficiency $\eta$ can be written as $V^{'}_{\pm} = \eta V_{\pm} + (1 - \eta)$. The total detection efficiency measured in our experiment is $44\%\pm 2\%$, corresponding to about 56\% loss, in good agreement with independent measurements of the loss sources: mode mismatch between the squeezed beam and the OMC cavity ($25\% \pm 5\%$),  scatter and absorption in the OMC ($18\% \pm 2\%$), and absorption and imperfect polarization alignment in the Faraday isolators (the squeezed beam passes through a Faraday three times, with total losses of $20\% \pm 2\%$). A significant reduction of these losses is possible, but it couldn't be achieved on the time scale allowed for this experiment before the H1 upgrade for Advanced LIGO began.
With 10.3 dB of squeezing leaving the OPO and a detection efficiency $\eta = 0.44$, only 2.2 dB of squeezing can be measured. 

We must also account for the impact of phase noise. Assuming that the relative phase noise between the local oscillator and the two squeezing quadratures has a normal distribution with a small standard deviation of $\tilde{\theta}$, the detected squeezing quadratures can be written as $V^{''}_{\pm} = V^{'}_{\pm}\cos^2\tilde{\theta} +V^{'}_{\mp} \sin^2\tilde{\theta}$~\cite{phTakeno2007, FHDFS06, PhaseTrack}.
An independent measurement indicates a phase noise of $37\pm6$ mrad. 

The detectable squeezing in our experiment is therefore $2.14\pm0.13$ dB, consistent with the measured sensitivity improvement of $2.15\pm0.05$ dB shown in Fig.~\ref{fig:sens20W}. Even though the impact of phase noise is negligible in this case, a correct accounting of phase noise is crucial to predict the detectable squeezing for higher detection efficiency and higher squeezing levels. For example, with 35 mrad of phase noise, a pure squeezed state of 20 dB injected into an interferometer with perfect detection efficiency would result in less than 9 dB of squeezing. 

A significant upgrade, known as Advanced LIGO~\cite{aLIGO2010}, is currently underway with the goal of increasing the strain sensitivity of the LIGO detectors by a factor of 10. An important element of the improved sensitivity is the increased light power. The nearly 1 MW of light power circulating inside the arms is close to the limits of the instrument, due to thermal effects caused by light absorption and the potential for optomechanical parametric instabilities~\cite{ParametricInstabilities}. Any further improvement looks best to be done with quantum-enhancing techniques~\cite{SqzReview2010, SqzReview2011}, such as squeezed vacuum injection. 

\begin{figure}
\centering
\includegraphics[width=0.48\textwidth]{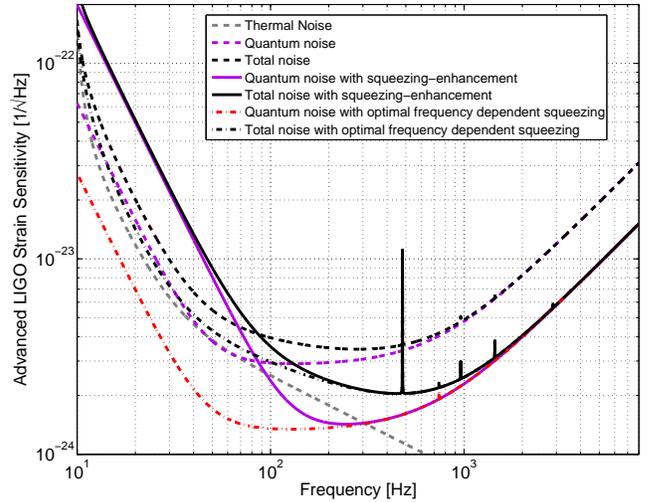}
\caption{Comparison of possible sensitivity curves for Advanced LIGO. Projection for a squeezing-enhanced Advanced LIGO interferometer
(continuous lines), using a design similar to the one described in this
paper, is compared to the Advanced LIGO sensitivity tuned for high
frequency performance (dashed lines). The total noise, in both cases, has been computed
by considering all the main noise sources, but only thermal noise
and quantum noise are shown, as they are the only relevant noise sources
above 100 Hz. The total losses for the squeezed beam were assumed to be
10\%, starting with 9 dB of squeezing delivered by the OPO and 35 mrad of phase noise. With the same parameters, but assuming the injection of optimal frequency dependent squeezing, quantum noise can be reduced at all frequencies as shown by the dash-dotted red line. \label{fig:quantumLIGO}}
\end{figure}

Fig.~\ref{fig:quantumLIGO} shows how Advanced LIGO could benefit from squeezing. The total losses for a squeezed beam in Advanced LIGO are expected to be significantly less than the losses measured in our experiment, possibly as low as 10\%. With the squeezed vacuum source employed in the H1 experiment, we could expect to reduce the shot noise by at least a factor of 2, improving the high frequency sensitivity of Advanced LIGO. Due to effective mitigation of other low-frequency noise sources, radiation pressure noise will now limit the low frequency sensitivity of Advanced LIGO. Thus, without further manipulation of the injected field~\cite{KLMTV2001}, the quantum enhancement at high frequencies would be achieved at the expense of the low frequency performance.
That said, a factor of 2 improvement, even at high frequencies, would significantly impact the astrophysical reach of the Advanced LIGO detectors for several types of sources. For, example, there are several dozen known pulsars with expected emission frequencies between 80 Hz and 776 Hz~\cite{S5CWallsky2012}. Probing the (quasi-)stationary quadrupole deformation of these spinning neutron stars will provide a constraint on the size of quadrupole deformations ("mountains") and thus, to some extent, on the breaking strain of the neutron star crust. Even more remarkable is that a factor of 2 increase of the high-frequency sensitivity will allow us to track the late inspiral and eventual merger of neutron star and black hole binary systems to higher frequencies. This will impose much more significant constraints on the poorly understood nuclear equation of state by measurements of the tidal deformability~\cite{tidal2010, tidal2011}, of the post-merger survival time~\cite{merger2012}, and of mode pulsations of the merged remnant (at frequencies $>$ 2 kHz).

Advanced LIGO can be improved with squeezing at all frequencies, as shown in Fig.~\ref{fig:quantumLIGO}, by arranging to inject squeezed vacuum with different quadrature angles at different frequencies (frequency-dependent squeezing~\cite{KLMTV2001}). We are currently developing the techniques for converting squeezed vacuum of the type produced here into frequency dependent squeezed vacuum for use in Advanced LIGO. By further improving the sensitivity of ground-based detectors and pushing the limits of astrophysical observations, quantum-enhancement techniques promise to play a critical role in future discoveries of gravitational wave sources. 

We have shown previously~\cite{corbitt2009} that the LIGO optomechanical system can be treated as a quantum oscillator with an occupancy number around 200. Given the extreme fragility of quantum mechanical states, it is all the more difficult to quantum engineer them without doing harm. With the result presented here, we have demonstrated that we can improve the sensitivity of a macroscopic quantum instrument without penalty.  This is of great relevance for Advanced detectors as they are expected to operate close to the quantum ground state of the optomechanical system.  

\section{Methods}

\subsection{Injection of squeezed vacuum}
A schematic of the squeezed vacuum source is shown in the grey box of Fig.~\ref{fig:layout}. The 1064 nm ``pump laser'' is phase-locked to the ``H1 laser'' and it drives the second harmonic generator (SHG) to produce light at 532 nm. 
The optical parametric oscillator (OPO) is resonant for both 1064 nm and 532 nm light. It is typically pumped with about 40 mW of 532 nm light, where the threshold for spontaneous sub-harmonic generation is near 95 mW. 
The 1064 nm `control laser'' is phase-locked to the ``pump laser" to generate a 29 MHz frequency shifted coherent beam which co-propagates with the squeezed vacuum beam, entering the interferometer through the ``output Faraday isolator''. The interferometer reflects both fields back towards the output mode cleaner (OMC). A 1\% sample of these two beams is detected before reaching the OMC by the ``squeezing angle control photodiode.'' 
The beat between the 29 MHz frequency shifted coherent beam and the interferometer beam provides an error signal which is used to control the phase of the squeezed vacuum field relative to the interferometer field.

\subsection{Optical Losses}
The optical losses measured in the path from the squeezed vacuum source to the ``output photodiode" are 56\%.  The dominant loss sources are: mode mismatch between the squeezed beam and the OMC cavity ($25\% \pm 5\%$),  scatter and absorption in the OMC ($18\% \pm 2\%$), and absorption and imperfect polarization alignment in the Faraday isolators (with total losses of $20\% \pm 2\%$). 
The mode mismatch between the squeezed beam and the output mode cleaner (OMC) is mainly caused by a complicated optical train in the vacuum envelope, which precluded improving the mode matching on a time scale compatible with this experiment. The losses in the OMC itself are also larger than expected, and they are believed to be due to scatter and absorption inside the mode cleaner cavity. 
The squeezed beam had to pass through a Faraday isolator installed between the squeezed vacuum source and the interferometer, and it had to double pass the ``output Faraday isolator.'' The large beam size out of the interferometer required us to use large aperture Faraday isolators. Large aperture Faraday isolators tend to have lower throughput due to the requirement for larger crystals. 
Most of these losses are due to the fact that the LIGO H1 detector was not initially designed for injection of squeezed states, and the squeezing injection path was retrofitted within the original LIGO optical layout.

\section{Acknowledgment} The authors gratefully acknowledge the support of the United States
National Science Foundation for the construction and operation of the
LIGO Laboratory and the Science and Technology Facilities Council of the
United Kingdom, the Max-Planck-Society, and the State of
Niedersachsen/Germany for support of the construction and operation of
the GEO600 detector. The authors also gratefully acknowledge the support
of the research by these agencies and by the Australian Research Council,
the International Science Linkages program of the Commonwealth of Australia,
the Council of Scientific and Industrial Research of India, the Istituto
Nazionale di Fisica Nucleare of Italy, 
the Spanish Ministerio de Econom\'ia y Competitividad,
the Conselleria d'Economia, Hisenda i Innovaci\'o of
the Govern de les Illes Balears, the Royal Society, the Scottish Funding 
Council, the Scottish Universities Physics Alliance, 
The National Aeronautics and Space Administration, 
the National Research Foundation of Korea,
Industry Canada and the Province of Ontario through the Ministry of Economic Development and Innovation, 
the National Science and Engineering Research Council Canada,
the Carnegie Trust, the Leverhulme Trust, the David
and Lucile Packard Foundation, the Research Corporation, and the Alfred
P. Sloan Foundation.
\section{Author Contribution} The activities of the LIGO Scientific Collaboration (LSC) cover modeling astrophysical sources of gravitational waves, setting sensitivity requirements for observatories, designing, building and running observatories, carrying out research and development of new techniques to increase the sensitivity of these observatories, and performing searches for astrophysical signals contained in the data. S. Dwyer,  S. Chua, L. Barsotti and D. Sigg were the leading scientists on this experiment, but a number of LSC members contributed directly to its success. M. Stefszky, A. Khalaidovski, M. Factourovich and C. Mow-Lowry assisted with the development of the squeezed vacuum source under the leadership of N. Mavalvala, D. McClelland and R. Schnabel. K. Kawabe supervised the integration of the squeezed vacuum source into the LIGO interferometer, with invaluable support from M. Landry and the LIGO Hanford Observatory staff. N. Smith-Lefebvre, M. Evans, R. Schofield and C. Vorvick kept the LIGO interferometer at its peak sensitivity and supported the integration of the squeezed vacuum source, with contributions from G. Meadors and D. Gustafson. The initial manuscript was written by L. Barsotti, N. Mavalvala, D. Sigg and D. McClelland. The LSC review of the manuscript was organized by S. Whitcomb. All authors approved the final version of the manuscript.
\section{Competing Interests} The authors declare that they have no
competing financial interests.
\section{Correspondence}Correspondence and requests for materials
should be addressed to Lisa Barsotti (lisabar@ligo.mit.edu).

\end{document}